\documentclass[preprint2]{emulateapj}
\usepackage{graphicx}
\usepackage{mathtools}
\usepackage{multirow}
\usepackage{epstopdf} 
\def\hi{{\mbox{\sc Hi}}}
\def\hii{{\mbox{\sc Hii}}}

\shorttitle{}
\shortauthors{}

\begin{document}

\title{The Multi-phase Turbulence Density Power Spectra in the Perseus Molecular Cloud}

\author{N.M., Pingel,\altaffilmark{1,2} Min-Young Lee,\altaffilmark{3,4} Blakesley Burkhart\altaffilmark{5}, Snezana Stanimirovi{\'c}\altaffilmark{6}}

\altaffiltext{1}{Department of Physics and Astronomy, West Virginia University,
White Hall, Box 6315, Morgantown, WV 26506; nipingel@mix.wvu.edu}
\altaffiltext{2}{Center for Gravitational Waves and Cosmology, West Virginia University, 
Chestnut Ridge Research Building, Morgantown, WV 26505}
\altaffiltext{3}{Max-Planck-Institut f\"ur Radioastronomie, 
Auf dem H\"ugel 69, 53121 Bonn, Germany}
\altaffiltext{4}{Laboratiore, AIM, CEA/IRFU/Service d' Astrophysique, 
Bat 709, 91191 Gif-sur-Yvette, France}
\altaffiltext{5}{Harvard-Smithsonian Center for Astrophysics, 60 Garden St., Cambridge, MA 0213}
\altaffiltext{6}{Department of Astronomy, University of Wisconsin, Madison, WI 53706}

\begin{abstract}
We derive two-dimensional spatial power spectra of four distinct interstellar medium tracers, $\hi$, $^{12}$CO($J$=1--0), $^{13}$CO($J$=1--0), and dust, in the Perseus molecular cloud, covering linear scales ranging from $\sim$0.1 pc to $\sim$90 pc. Among the four tracers, we find the steepest slopes of $-3.23\pm0.05$ and $-3.22\pm0.05$ for the uncorrected and opacity-corrected $\hi$ column density images. This result suggests that the $\hi$ in and around Perseus traces a non-gravitating, transonic medium on average, with a negligible effect from opacity. On the other hand, we measure the shallowest slope of $-2.72\pm0.12$ for the 2MASS dust extinction data and interpret this as the signature of a self-gravitating, supersonic medium. Possible variations in the dust-to-gas ratio likely do not change our conclusion. Finally, we derive slopes of $-3.08\pm0.08$ and $-2.88\pm0.07$ for the ${}^{12}$CO(1--0) and ${}^{13}$CO(1--0) integrated intensity images. Based on theoretical predictions for an optically thick medium, we interpret these slopes of roughly $-3$ as implying that both CO lines are susceptible to the opacity effect. While simple tests for the impact of CO formation and depletion indicate that the measured slopes of ${}^{12}$CO(1--0) and ${}^{13}$CO(1--0) are not likely affected by these chemical effects, our results generally suggest that chemically more complex and/or fully optically thick media may not be a reliable observational tracer for characterizing turbulence. 

\end{abstract}

\keywords{: ISM: clouds --- ISM: structure --- magnetohydrodynamics (MHD) – turbulence}

\section{Introduction}
\label{s:intro}
\setcounter{footnote}{0}

Turbulence has been considered as one of the key physical processes for the evolution of the interstellar medium (ISM) (e.g., \citealt{ElmeScalo04, McKeeOst07,laz09}). Yet many questions still remain open, in particular with regard to the role of turbulence in the formation and evolution of molecular clouds (MCs) and subsequent star formation. Some of the important questions include: What are the injection sources of turbulent energy and on what scales? How is the transition from atomic ($\hi$) to molecular hydrogen (H$_{2}$; \citealt{BialyBurkhartSternberg17}), likely a prerequisite of star formation, influenced by turbulence? How does turbulence manifest itself as a function of ISM phase (e.g., cold and warm neutral medium (CNM and WNM) and molecular gas)? How does opacity affect an observer's ability to accurately characterize the properties of turbulence?  

As for the origin of turbulence, it is believed that turbulence is driven on a variety of spatial scales and cascades down to smaller scales, as evidenced by the the fractal structure of the ISM (e.g., \citealt{stutzki98, stan99, ElmeKim01}). The accretion of the circumgalactic medium and galactic-scale gravitational instabilities likely trigger turbulence on large scales (e.g., \citealt{kless10, krumBurk16}, while stellar feedback such as outflows and supernova explosions inject energy on smaller scales (e.g., \citealt{krumMatznerMcKee06, zamora12,padoan16}). 

Depending on the specific driver, the characteristics of turbulence will then be imprinted within the ISM mainly as three-dimensional density and velocity fluctuations, and these fluctuations have been traditionally studied via correlation functions such as the spatial power spectrum (SPS) (e.g., \citealt{crovDickey83}), $\Delta$-variance (e.g.~\citealt{stutzki98}, and structure function (e.g., \citealt{padCamLang02, burkhart15_SF}). In particular, the SPS approach has been applied to observations of various Galactic and extragalactic environments (e.g., \citealt{Plume00, Dickey01, ElmeKim01, Burkhart2010, Combes12, Zhang12, Pingel13}), showing power spectral slopes $\beta$ roughly ranging from $-2.7$ to $-3.7$ depending on the used tracers (e.g., $\hi$, carbon monoxide (CO), and dust). These slopes essentially provide information on the relative amount of structures as a function of spatial scale and can be compared with theoretical models of turbulence (mainly numerical simulations) to characterize turbulence cascade (e.g., \citealt{Burkhart2010}), to determine the influence of shocks (e.g., \citealt{Beresnyak05}), to reveal the injection and dissipation scales of turbulent energy (e.g., \citealt{KowalLazarian07, FederrathKlessen13,Chen15}), and to trace the evolution of MCs (e.g., \citealt{BurkhartCollins15}). 
The proximity and a wealth of multi-wavelength observations make MCs in the solar neighborhood an ideal laboratory for probing the impact of turbulence on the formation and evolution of MCs. In this paper, we focus on the Perseus molecular cloud, which is a nearby ($\sim$300 pc; e.g.~\citealt{HerbigJones83, Cernis90}), low-mass ($\sim$2 $\times$ 10$^{4}$ M$_{\odot}$; e.g.~\citealt{Sancisi74, Lada10}) cloud. Its star formation activities, as well as atomic and molecular gas contents, have been extensively examined over the past decade (e.g., \citealt{Ridge06, Jorgensen07, Pineda08, Lee12, Lee14, Lee15, Mercimek17}), revealing that the cloud consists of several individual dark and star-forming regions (e.g., B5, B1, B1E, IC348, and NGC1333) and is actively forming low- to intermediate-mass stars (see \citealt{Bally08} for a review). 

The properties of turbulence in Perseus has been studied in the past, mainly analyzing the observations of CO isotopologues ($^{12}$CO($J$=1--0, 3--2) and $^{13}$CO($J$=1--0, 2--1)\footnote{In this paper, we quote the $^{12}$CO($J$=1--0) and $^{13}$CO($J$=1--0) lines as $^{12}$CO and $^{13}$CO, while specifying other CO transitions.}; e.g., \citealt{Padoan99, Bensch01, Sun06}) and dust (e.g., \citealt{Schneider11, Burkhart15_PDF}) with the SPS, probability distribution function, and $\Delta$-variance methods. 
 
In this paper, we extend these previous studies, with an aim of probing the characteristics of turbulence in the multi-phase ISM more comprehensively. To this end, we perform a SPS analysis of multi-wavelength data, including recent high resolution ($\sim$4.3$'$ or $\sim$0.4 pc scales) $\hi$ observations of Perseus (\citealt{Lee12, Stan14, Lee15}), and systematically compare the derived power spectra. In addition, we examine several biases that could affect SPS analyses, such as the impact of opacity and CO chemistry, and discuss the optimal tracer(s) for studying the turbulent environments of MCs. 

Our paper is structured as follows. In Section~\ref{sec:data}, we describe the observations of each tracer used in this study. In Section~\ref{sec:analysis}, we then summarize the SPS technique and derivation of uncertainties.  Section~\ref{sec:results} presents the derived power spectra, and Section~\ref{sec:discussion} discusses how our results relate to the theoretical predictions of the SPS, as well as previous observational studies. Finally, in Section~\ref{sec:conclusion}, we summarize our conclusions. 

\section{Data} 
\label{sec:data}

In this section, we describe the gas and dust data used in our analysis of the SPS. Several specifics of the data, e.g., image/pixel sizes, angular resolutions, and median 1$\sigma$ uncertainties, are summarized in Table \ref{t:data}. 

\subsection{HI}
\label{s:HI}

We use the $\hi$ column density image of Perseus from \citet{Lee12}. To derive the $\hi$ column density on 4.3$'$ scales ($\sim$0.4 pc at the distance of Perseus), \citet{Lee12} used $\hi$ cubes from the GALFA-$\hi$ survey\footnote{https://purcell.ssl.berkeley.edu/} \citep{Peek11} and integrated the $\hi$ emission from $V_{\rm LSR}$ = $-$5 to $+$15 km s$^{-1}$ under the optically thin assumption. This velocity range was determined based on the correlation between the derived $N$($\hi$) and the 2MASS $A_{V}$ from the COMPLETE survey\footnote{https://www.cfa.harvard.edu/COMPLETE/} \citep{Ridge06}. The final $\hi$ column density image is centered at (R.A., decl.) = (03$^{\rm h}$29$^{\rm m}$52$^{\rm s}$,$+$30$^{\circ}$34$'$01$''$) with a size of $\sim$14.7$^{\circ}$ $\times$ 9.0$^{\circ}$. 

\begin{figure*} 
\centering{
\includegraphics[width=3.2in]{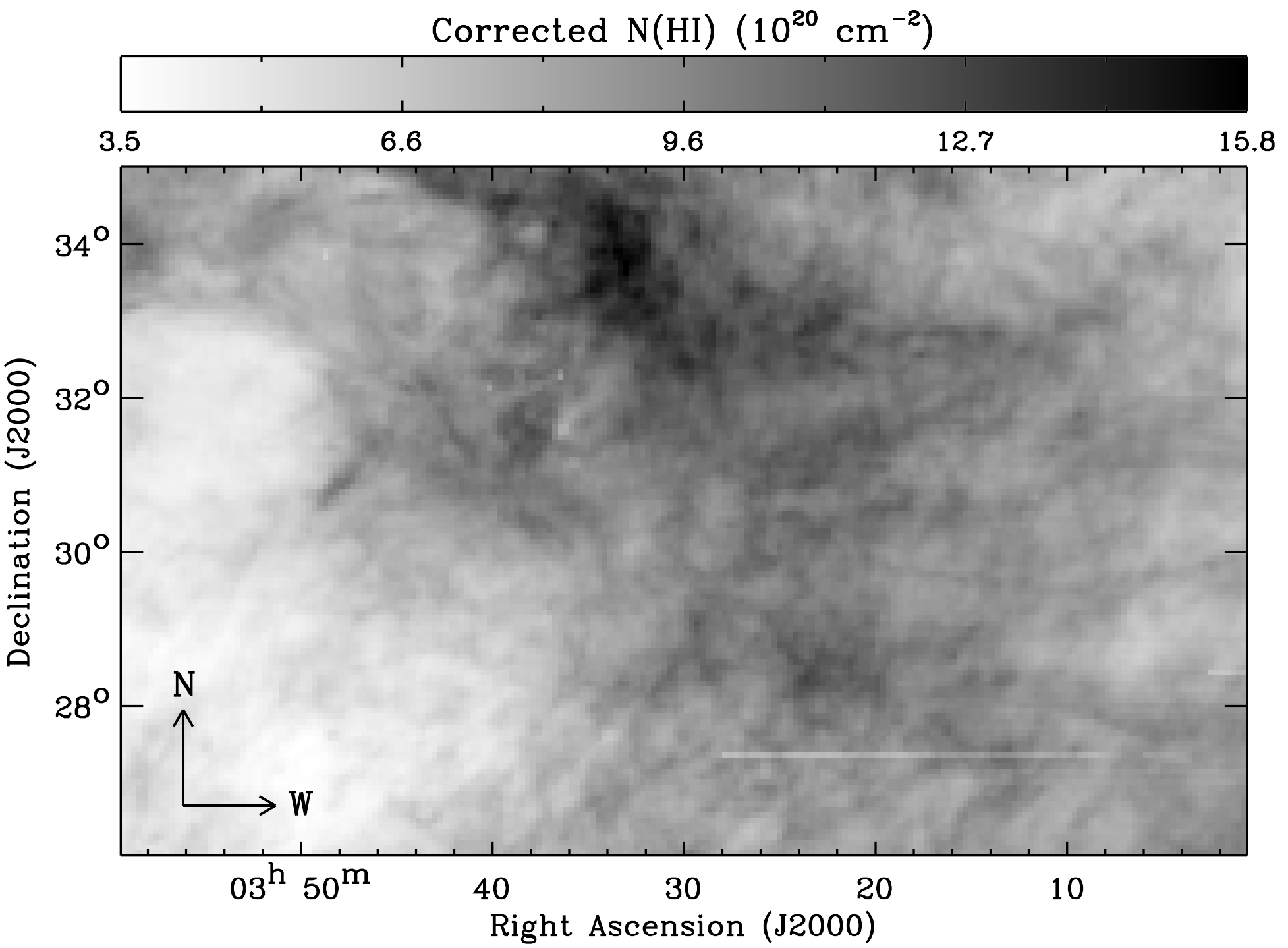} 
\includegraphics[scale=.4]{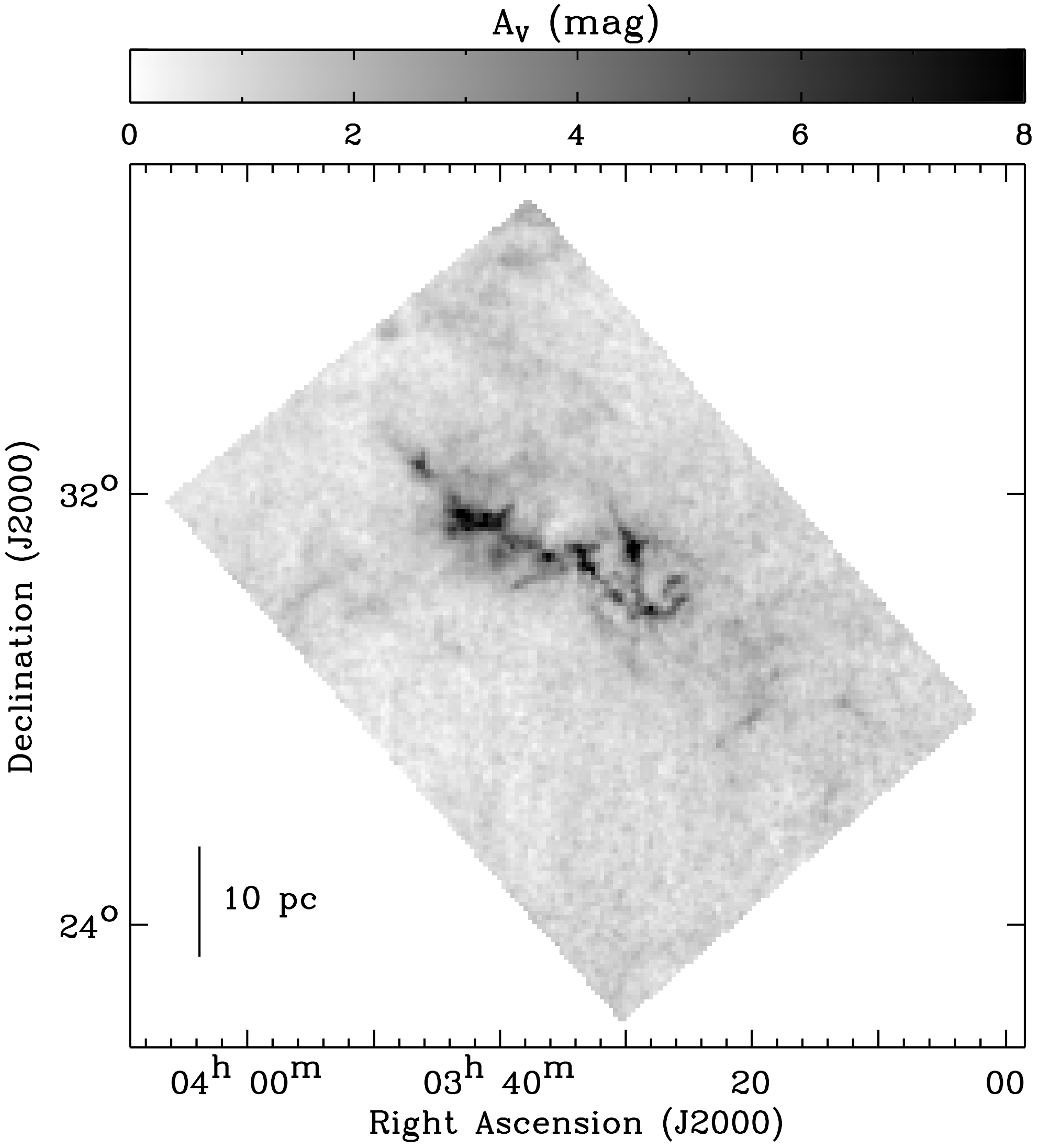}
\includegraphics[width=3.2in]{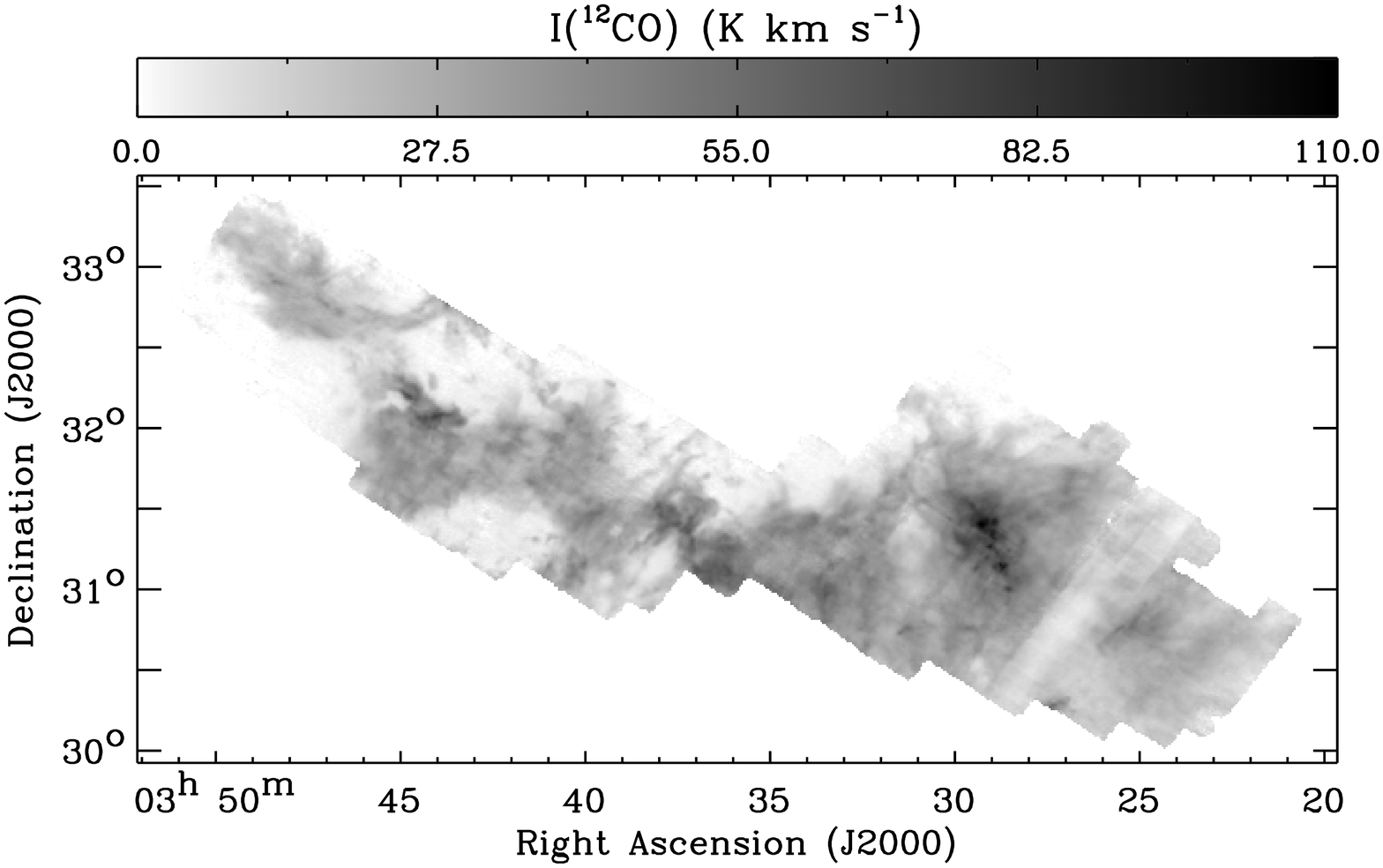}
\includegraphics[width=3.2in]{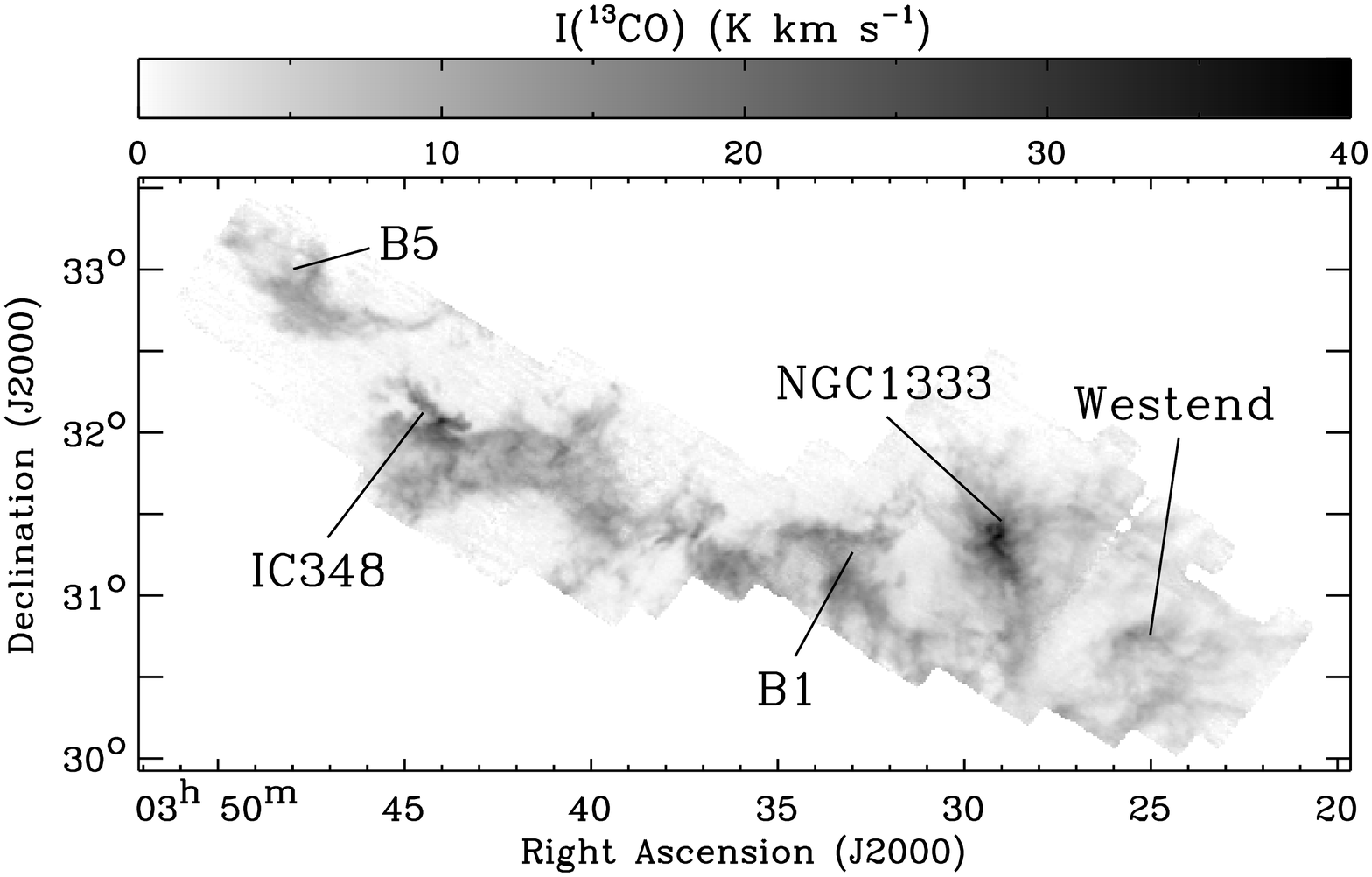}
\caption{\label{fig:data} Data used in our study. See Section \ref{sec:data} for details. 
(\textit{top/left}) Corrected HI column density. 
(\textit{top/right}) 2MASS $A_{V}$. 
(\textit{bottom/left}) FCRAO $^{12}$CO integrated intensity. 
(\textit{bottom/right}) FCRAO $^{13}$CO integrated intensity.
The five sub-regions used by Pineda, Caselli, \& Goodman et al. (2008) are labeled here (see Table \ref{tab:sub_regions}).}}
\end{figure*}

In addition, we make use of the $\hi$ column density image from \citet{Lee15}, which was corrected for optical depth effects, with an aim of probing the impact of optically thick $\hi$ on the slope of a power spectrum. In essence, \citet{Lee15} derived the empirical correction factor for the optically thick $\hi$ in Perseus based on Arecibo $\hi$ absorption measurements toward 26 background continuum sources \citep{Stan14} and applied the correction to the $\hi$ column density image from \citet{Lee12} on a pixel-by-pixel basis. The amount of the cold $\hi$ was found not substantial (median cold-to-total $\hi$ ratio of $\sim$0.3) and the opacity correction was hence small (only up to $\sim$1.2). The final opacity-corrected $\hi$ column density image is presented in Figure \ref{fig:data}, and we refer to \citet{Lee12} for details on the derivation of the $\hi$ column density images. 

\begin{table*} 
\begin{center} 
\caption{\label{t:data} Characteristics of the data used in our study}
\begin{tabular}{l c c c c c} \hline \hline 
 & $N$($\hi$) & Corrected $N$($\hi$) & $A_{V}$ & $I$($^{12}$CO(1--0)) & $I$($^{13}$CO(1--0)) \\ \hline 
\multirow{2}{*}{Image Size$^{\rm a}$} & $14.8^{\circ}\times9.0^{\circ}$ & $14.8^{\circ}\times9.0^{\circ}$ & $15.1^{\circ}\times15.5^{\circ}$ & $6.6^{\circ}\times3.5^{\circ}$ & $6.6^{\circ}\times3.5^{\circ}$ \\
                                      & (77.5 pc $\times$ 47.1 pc) & (77.5 pc $\times$ 47.1 pc) & (79.1 pc $\times$ 81.2 pc) & (34.6 pc $\times$ 18.3 pc) & (34.6 pc $\times$ 18.3 pc) \\ 
\multirow{2}{*}{Pixel Size$^{\rm a}$} & 4.3$'$ & 4.3$'$ & 5.0$'$ & 46.0$''$ & 46.0$''$ \\ 
                                      & (0.4 pc) & (0.4 pc) & (0.4 pc) & (0.07 pc) &  (0.07 pc) \\ 
Angular Resolution & 4.3$'$ & 4.3$'$ & 5.0$'$ & 46.0$''$ & 46.0$''$ \\ 
Median 1$\sigma$ & 6 $\times$ 10$^{19}$ cm$^{-2}$ & 8 $\times$ 10$^{19}$ cm$^{-2}$ & 0.2 mag & 0.5 K km s$^{-1}$ & 0.2 K km s$^{-1}$ \\
\hline
\end{tabular}
\end{center}
{$^{a}$ The distance to Perseus is assumed to be 300 pc.} 
\end{table*}

\subsection{$A_{V}$}
\label{s:Av}

We use the $A_{V}$ image from the COMPLETE survey, which was produced by applying the NICER technique \citep{LombardiAlves01} to 2MASS data\footnote{http://www.ipac.caltech.edu/2mass/releases/allsky/doc/explsup.html} at 5$'$ resolution. For our analysis, we regrid the $A_{V}$ image to have independent pixels with a size of 5$'$ (Figure \ref{fig:data}). Note that \citet{Lee12} also derived the $A_{V}$ image of Perseus based on 60 $\mu$m and 100 $\mu$m data from the IRIS survey \citep{Miville05}. We, however, do not use this image due to the presence of masked pixels, which will cause the ``Gibbs ringing'' in our power spectrum analysis (see Section \ref{sec:analysis} for details). In \citet{Lee12}, $\sim$20\% of the total pixels were blanked to exclude the lines of sight where dust emission modeling was unreliable due to possible contaminations from foreground stars, a background $\hii$ region, and the nearby Taurus molecular cloud. A pixel-by-pixel comparison shows that the \citet{Lee12} $A_{V}$ and the 2MASS $A_{V}$ are in agreement: they have overall a 1:1 relation with a small median difference of $\sim$0.2 mag (which is comparable to the median 1$\sigma$ uncertainty of the 2MASS $A_{V}$ image; Table \ref{t:data}). This agreement primarily arises from the fact that \citet{Lee12} estimated $A_{V}$ by calibrating the derived dust optical depth at 100 $\mu$m with the 2MASS $A_{V}$. The scatter around the 1:1 relation does exist, however, and this can be mostly attributed to the variations in the dust temperature along a line of sight \citep{Goodman09}. 

\subsection{CO} 
\label{subsec:CO}

Finally, we use the $^{12}$CO($J$=1--0) and $^{13}$CO($J$=1--0) cubes from the COMPLETE survey \citep{Ridge06}. These cubes were obtained with the 14-m FCRAO telescope at 46$''$ resolution ($\sim$0.07 pc at the distance of Perseus), covering an area of $\sim$6.6$^{\circ}$ $\times$ 3.5$^{\circ}$ with a center of (R.A.,decl.) = (03$^{\rm h}$36$^{\rm m}$03$^{\rm s}$,$+$31$^{\circ}$44$'$10$''$). To derive the $^{12}$CO and $^{13}$CO integrated intensity images (Figure \ref{fig:data}), we first regrid the cubes to have a pixel size of 46$''$ and integrate the CO emission from $V_{\rm LSR}$ = $-$5 to $+$15 km s$^{-1}$. 

\section{SPS Analysis}
\label{sec:analysis}
\subsection{Derivation of the SPS}
\label{subsec:SPSDerivation}
\begin{table}
\caption{Summary of Moduli Properties} 
\label{tab:modStats}
\centering{\resizebox{\columnwidth}{!}{\begin{tabular}{ccccc}
\hline \hline
\\[-1.0em]
\\[-1.0em]
  & $\hi$ & $A_{V}$ & ${}^{12}$CO(1--0) & ${}^{13}$CO(1--0) \\
\\[-1.0em]
\hline
\\[-1.0em]
\\[-1.0em]
Grid Size [pixels] & 206$\times$126 & 200$\times$200 & 520$\times$273 & 520$\times$273 \\
Angular Extent [deg] & 14.8 & 16.7 & 6.6 & 6.6 \\
Modulus Pixel Resolution [$\lambda$] & 3.9 & 3.4 & 8.6 & 8.6 \\
Longest Spatial Frequency [$\lambda$] & 397.6 & 341.0 & 2241.0 & 2241.0 \\ 
Smallest Effective Baseline [$\lambda$] & 10.1 & 8.9 & 27.5 & 27.5 \\
Maximum Linear Scale [pc] & 77.7 & 87.9 &  34.8 & 34.8 \\ 
Minimum Linear Scape [pc] & 0.8 & 0.9 & 0.1 & 0.1 \\
Maximum Effective Linear Scale [pc] & 29.6 & 33.7 & 10.9 & 10.9 \\ \hline 
\end{tabular}}}
\end{table}
The SPS, i.e.~the Fourier transform of the two-point auto-correlation function, provides information on the properties of turbulence, including the injection and dissipation scales, as well as inertial range scaling. The two-dimensional SPS is defined as:
\begin{equation}
P(\nu)=\mathfrak{F}\left(\nu\right)\times \mathfrak{F}^{*}\left(\nu\right)
\label{eq:SPS}
\end{equation}

where $\nu$ is the wavenumber ($\nu=\frac{2\pi}{L}$ where $L=$ length scale) and $\mathfrak{F}\left(\nu\right)$ is the Fourier transform of the field under study, which for our purpose is the $\hi$/dust column density and $^{12}$CO/$^{13}$CO integrated intensity images of Perseus. 

Our derivation of the SPS is similar to previous studies (e.g.,~\citealt{crovDickey83,green93,stan99,Elmegreen01, Muller04, Pingel13}). 

In practice, we take the 2D Fourier transform of each column density or integrated intensity image and square the modulus of the transform --- $<$$\Re^2$+$\Im^2$$>$, where $\Re$ and $\Im$ are the real and imaginary components of the 2D Fourier transform, respectively --- in order to derive what we define as the modulus image with which we perform our statistical analysis. 

We relate each baseline to a linear length scale by adopting the distance to Perseus to be 300 pc. Specifically, the linear length scale, is related to the spatial frequency ($k$; measured in units of wavelength $\lambda$) by
\begin{equation}
L~[pc] \sim \frac{d~[pc]}{k~[\lambda]},
\label{eq:scale}
\end{equation} 
where $d$ is the distance to Perseus and $k$ represents the spatial frequency. For the $\hi$ data, the maximum angular scale of 14.8$^{\circ}$ corresponds to a linear length scale of 77.5 pc, which sets the pixel resolution of 3.9$\lambda$. The origin is set at the center of the Fourier plane, and the spatial frequencies range from 0 to 397.6$\lambda$. The grid size, pixel resolution, and spatial frequency range for the $\hi$ and other tracers are summarized in Tables~\ref{tab:modStats}. 

\begin{figure}
\includegraphics[width=3.5 in]{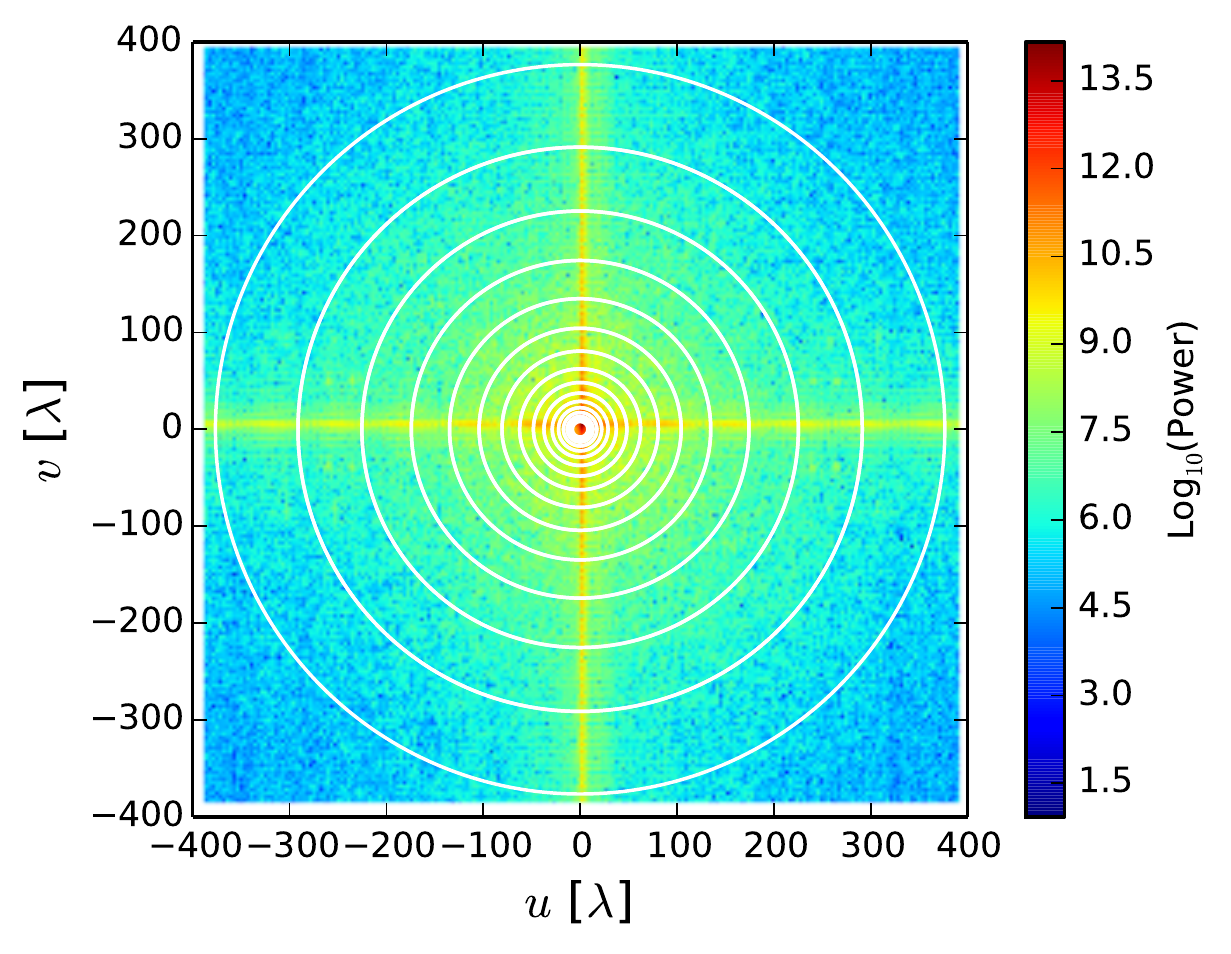}
\caption{\label{fig:modImage} Modulus image ($<$$\Re^2$+$\Im^2$$>$) derived from the uncorrected $\hi$ data. 
The very bright pixels along the axes are artifacts from the Gibbs phenomenon caused by the Fourier transform of the sharp image. 
The white contours show the boundaries of the annuli within which we perform our statistical analysis.}
\end{figure}

\begin{figure*}
\centering{
\includegraphics[width = 7.5 in]{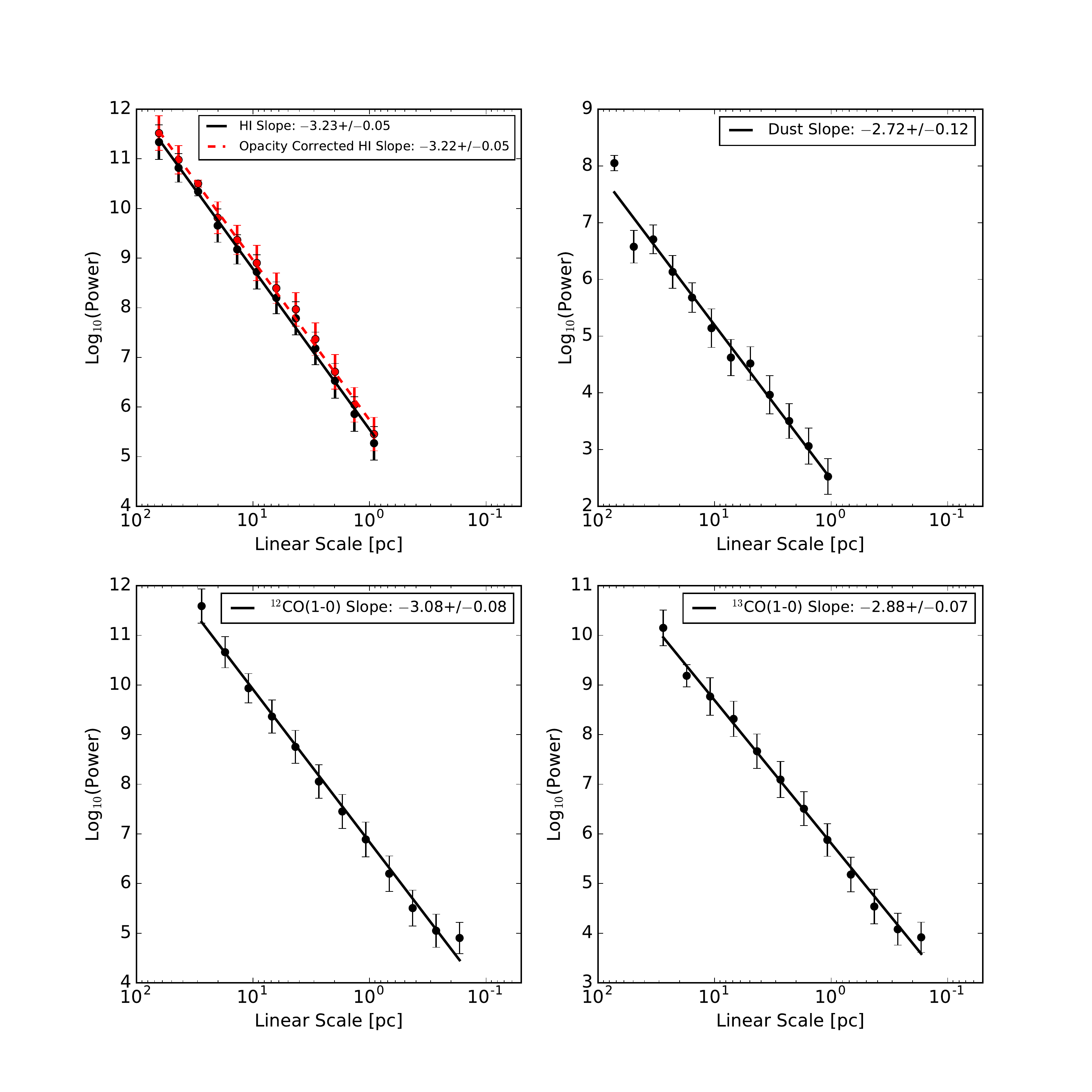}

\caption{Intensity map power spectra derived for our tracers. 
\textit{Top:} $\hi$ and $A_{V}$. 
The red and black points in the $\hi$ plot represent the power measured in the opacity-corrected and uncorrected images, respectively. 
\textit{Bottom:} ${}^{12}$CO(1--0) and ${}^{13}$CO(1--0).}
\label{fig:intPS}}
\end{figure*}

To derive the SPS, we first place twelve annuli (uniformly spaced in log space) on each modulus image such that the innermost annulus corresponds to the largest length scale, while the outermost annulus corresponds to the smallest length scale. In this case, the linear length scale ranges from 77.7 pc down to 0.8 pc for the $\hi$, 87.9 pc down to 0.9 pc for the $A_{V}$, and 34.8 pc down to 0.1 pc for both CO data sets. As an example, we present the modulus image of the $\hi$ with the annuli boundaries overlaid in Figure~\ref{fig:modImage}. Next, assuming azimuthal symmetry in the modulus image, we calculate the median value of the pixel distribution contained within each annulus and plot the median values as a function of decreasing linear scale. The derived SPS for each tracer is presented in Figure~\ref{fig:intPS}. 
 
Our use of the median values is motivated by the significant number of bright pixels along the axes of each modulus image caused by the Gibbs phenomenon. The Gibbs phenomenon causes $sin(x)/x$ (where $x$ is the pixel coordinate) ringing along the axes of the modulus image due to the Fourier transform of discontinuous image edges. The resulting ``cross'' in the modulus image introduces a high-end tail in the pixel distributions within individual annuli. Traditionally, a power spectrum analysis has been done using mean values (e.g., ~\citealt{stan99, Dickey01, Muller04}). We experimented with padding the images with zeros, as well as applying Gaussian tapers similar to the methods by \citet{Muller04} to smoothly reduce the intensity to zero at the image edges, but found that the pixel distributions, especially in the $\hi$ image, still possess a high-end tail of pixel values. 

\begin{figure*}
\centering{
\includegraphics[width=3.2 in]{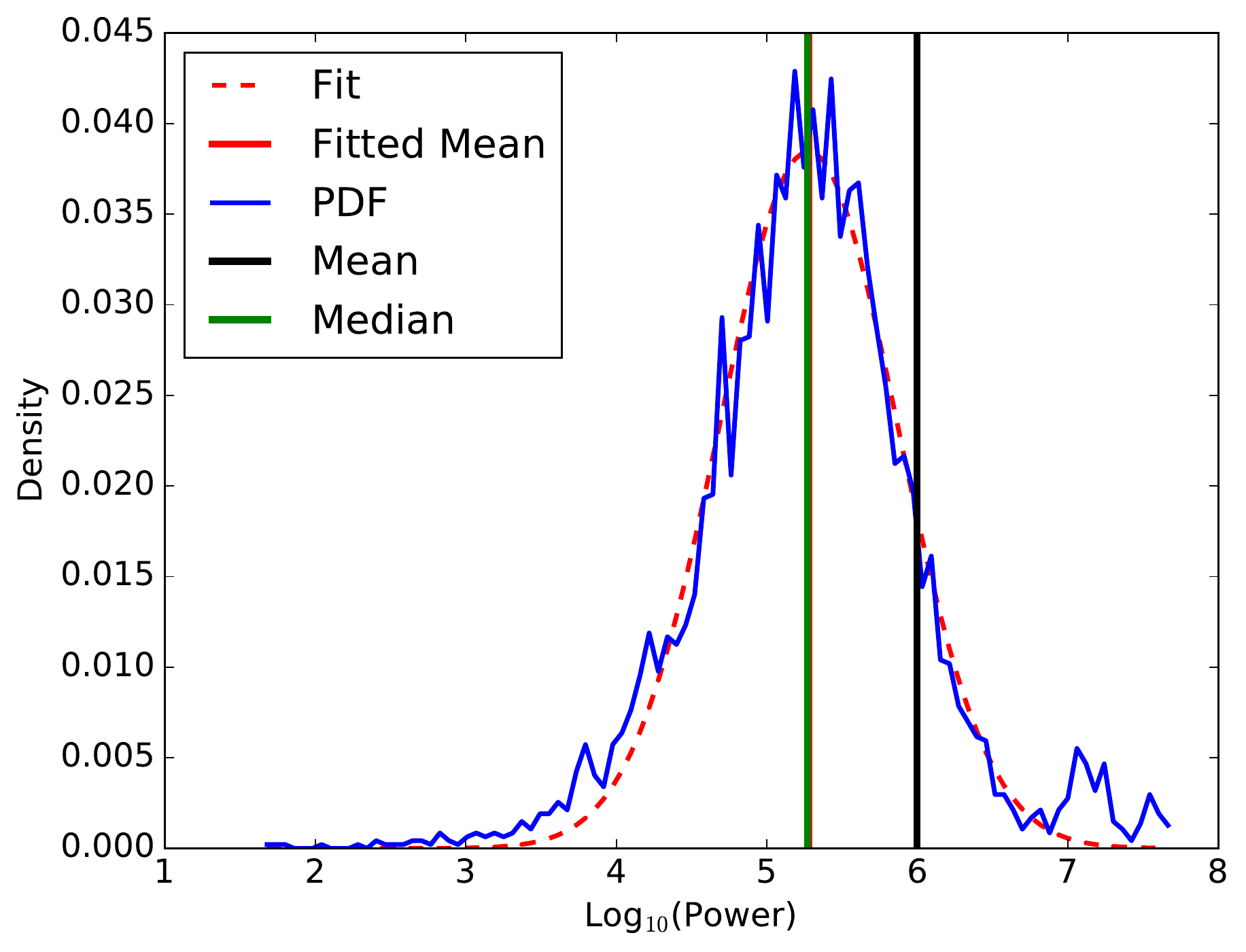}
\includegraphics[width=3.2 in]{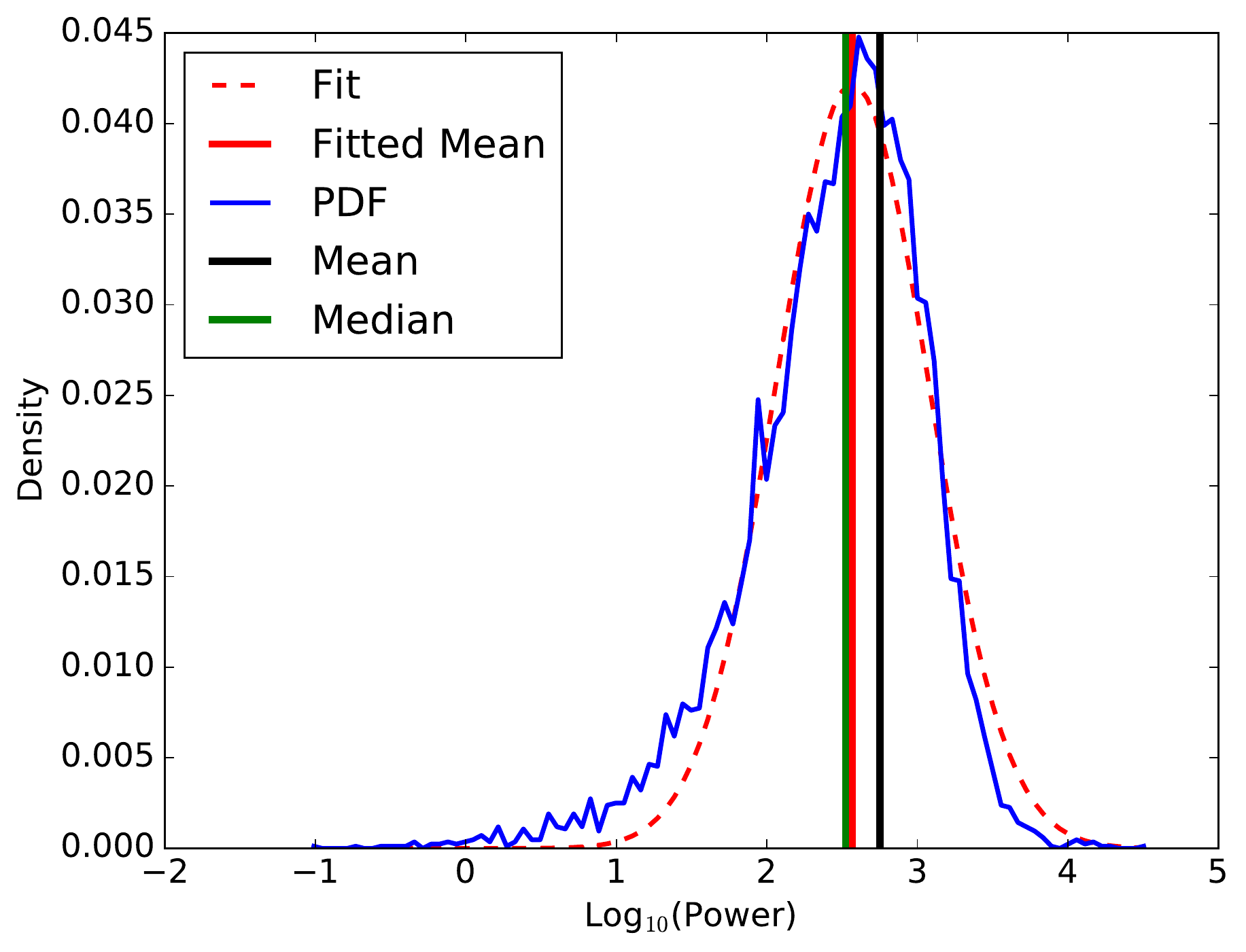}
\includegraphics[width=3.2in]{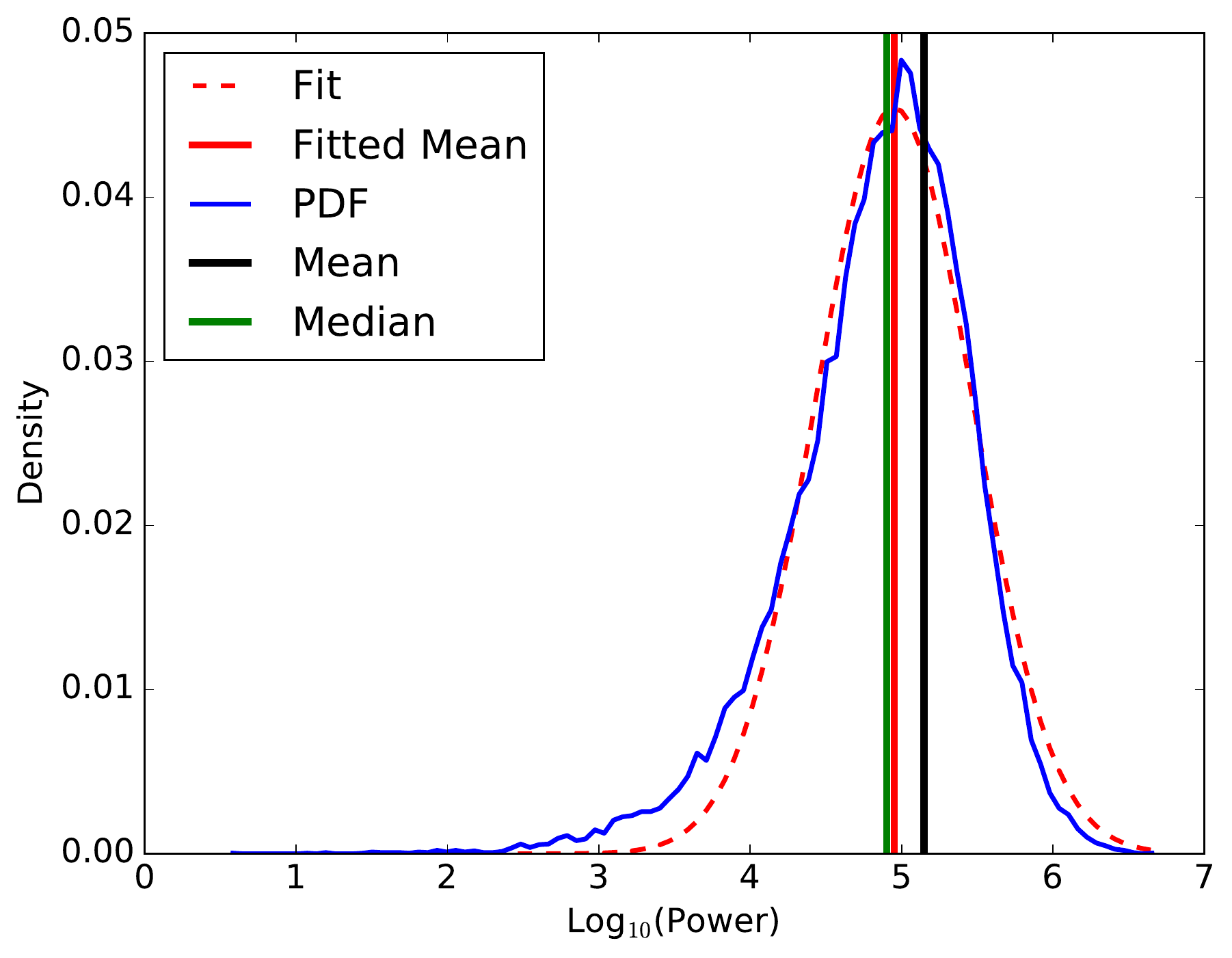}
\includegraphics[width=3.2 in]{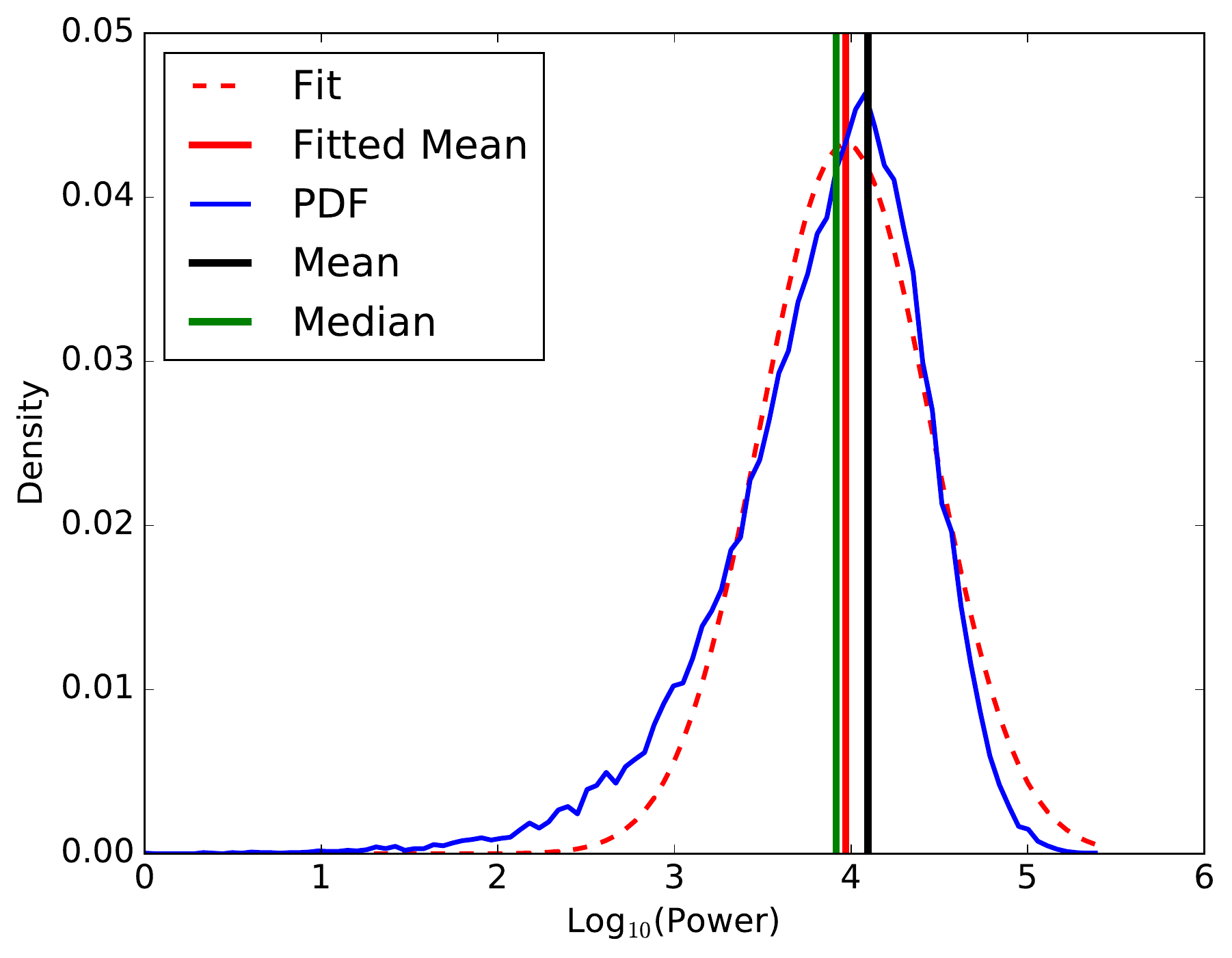}
\caption{\label{fig:pdf} Pixel distribution within the largest annulus (smallest length scale) for each tracer. The dashed red line represents a Gaussian fit to the distribution. 
The vertical red and black line represents the fitted and arithmetic mean of the distribution, respectively. The vertical green line shows the median value of the entire pixel distribution.
\textit{Top:} (Uncorrected) $\hi$ and $A_{V}$; \textit{Bottom:} {}$^{12}$CO(1--0) and ${}^{13}$CO(1--0).}}
\end{figure*}

To demonstrate the impact of the high-end tail, we present the probability distribution function (PDF) of the pixel distribution within the largest annulus for each tracer in Figure \ref{fig:pdf}. On average, we find that each pixel distribution can be characterized reasonably well by a Gaussian with some high-end tail. We thus fit a Gaussian function (red dashed line) to each PDF and overlay the arithmetic mean (black line), the fitted mean (red solid line), and the median (green line). For each tracer, the median of the PDF is consistently within $<$ 1\% of the fitted mean, suggesting that the median is a good representative value of each pixel distribution. The arithmetic mean, however, is consistently higher than what is expected from a Gaussian distribution, mainly due to the high power values created by the Gibbs phenomenon. The $\hi$ PDF clearly demonstrates the impact of these artificially high power values on the computed mean:  
only 2\% of the pixels have power values greater than 3$\sigma$ as determined by the Gaussian fit, but the arithmetic mean of the distribution is an order of magnitude higher than the median and mean returned by the fit. On the other hand, the mean values of the pixel distributions for other tracers are not as susceptible to the effects of the Gibbs phenomenon as $\hi$. 
The relatively well-behaved distributions for these tracers are due to the gradual decrease of integrated emission towards the edge of their respective images. The general uniformity in the $\hi$ integrated emission (the pixel values vary by only a factor of two over a large area of $\sim$15$^{\circ} \times 10^{\circ}$), and to some extent the dust tracer, enhances the edge effects of the Fourier transform, which in turn increases the discrepancy between the median and the arithmetic mean. Considering how well the median characterizes a pixel distribution even in the presence of outlying values, using the median as opposed to the mean of the pixel distributions of the modulus image is justified to derive power spectral slopes. We examine the PDFs of the pixel distributions in all individual annuli for each tracer and indeed confirm that the median is a robust representation of the distributions.  

Our final concern about the measured median power within each annulus is the contributions from noise and beams. We first find that the power spectra of $\hi$, {}$^{12}$CO(1--0), and ${}^{13}$CO(1--0) derived from emission-free channels only are flat (which is expected for white noise) and three or four orders of magnitude lower than those in Figure \ref{fig:intPS}. We therefore conclude that the contribution from noise to our derived power spectra is negligible. 

As for the effect of the beam, we expect no significant impact since the pixels in each of our tracer maps are independent with each other 
with the size equal to the respective beamsize. We confirm this by deriving the SPS of an idealized Gaussian beam with a FWHM set to the respective beamsize for each tracer and finding that the beam has a negligible contribution to the power at scales larger than the pixel resolution.  

\subsection{Derivation of SPS Uncertainties}
Following \citet{Pingel13}, we initially estimated the errors of the median values by (1) creating 1000 simulated integrated intensity or column density images (each pixel value of a simulated map was determined by drawing from a Gaussian distribution centered on the original map pixel value and width characterized by the associated 1$\sigma$ uncertainty), (2) deriving the SPS for each image, and (3) taking the final uncertainty for each annulus as the standard deviation of the measured 1000 median power values. For each tracer, we then found that this Monte Carlo approach results in an order of magnitude lower uncertainty on the largest length scale as compared to smaller length scales.

This discrepancy in uncertainties between different scales mainly arises from a broad difference in the number of pixels available for each annulus, i.e., the small number of pixels for the largest length scale tends to cause a small spread in individual simulated values, which drives the small standard deviation between the 1000 realizations. The lower uncertainties on large length scales in turn leads to artificially shallow slopes, since the corresponding data points are weighted more in our linear regression fit. 
 
We therefore require a method for error characterization that is robust against the difference in sample sizes, as well as high-end tail pixel values. To this end, we employ the median absolute deviation (MAD) method. The MAD of some distribution $X$, which depends on a univariate data set made up of individual samples ($X_1$,$X_2$,...$X_n$), is defined as
\begin{equation}
MAD = median\left(\lvert X_i - median\left(X\right)\rvert\right).
\end{equation}
In practice, we calculate the absolute deviations of the pixel values from the computed median for each annulus and use the median of these deviations as a final error. This MAD is more resilient against outlying values than standard deviation. For example, in the case of standard deviation, the large differences between pixel values and the artificially high mean due to the Gibbs ringing are squared causing these large discrepancies to be weighted more. In the MAD calculation, on the other hand, a small number of large absolute deviations becomes less relevant due to the use of the median.

\section{Results}
\label{sec:results}

The derived power spectra in Figure \ref{fig:intPS} show that the four tracers have a range of slopes, suggesting that turbulence manifests itself in different ways depending on the medium: the 2MASS $A_{V}$ ($-2.72\pm0.12$) shows the shallowest slope, while the $^{12}$CO(1--0) ($-3.08\pm0.08$), ${}^{13}$CO(1--0) ($-2.88\pm0.07$), and $\hi$ ($-3.23\pm0.05$ and $-3.22\pm0.05$ for the uncorrected and corrected $\hi$ respectively) show the intermediate and steepest slopes. Among these four tracers, ${}^{12}$CO(1--0), ${}^{13}$CO(1--0), and dust have been previously examined in the context of the turbulent environment of Perseus. One such study was \citet{Padoan06_a}. 
In their SPS analysis, the same ${}^{13}$CO(1--0) image from \citet{Ridge06} was used to derive the spectral slope ($\beta_I$) of $-1.99\pm0.05$ by measuring the \textit{total} power, rather than the \textit{average} or \textit{median} value, within each wavenumber shell. Since we take a similar derivation approach for the power spectral index as \citet{lazPogo00}, our slope ($\gamma$) is related to the estimate by \citet{Padoan06_a} in the following way: $\beta_I$ $=$ $1-\gamma$. In order to properly compare with our result, a value of one must then be subtracted from the \citet{Padoan06_a} slope. Considering other differences between the two studies (e.g., angular resolution of 46$''$ and 92$''$ for our study and \citet{Padoan06_a} respectively), we conclude that our ${}^{13}$CO(1--0) power spectral slope is consistent with \citet{Padoan06_a}

\citet{Sun06} is another study which utilized similar ${}^{12}$CO(1--0), ${}^{13}$CO(1--0), and 2MASS $A_{V}$ datasets to investigate the properties of turbulence in Perseus. To measure power spectral indices, they employed the $\Delta$-variance technique and found the slopes of $-3.08\pm0.04$, $-3.09\pm0.09$, and $-2.55\pm0.02$ for $^{12}$CO(1--0), $^{13}$CO(1--0), and dust respectively. 
Considering slight differences in the datasets, as well as the fitting ranges, we conclude that our results are in relatively good agreement with \citet{Sun06}.   

Interestingly, for all tracers, we do not find a break over two orders of magnitude in length scales ($\sim$0.1 pc to $\sim$10 pc). This implies that turbulence is most likely driven on the scales larger than the size of the entire molecular cloud and dissipated down to the scales smaller than $\sim$0.1 pc (see Section \ref{s:final_discussion} for more discussions). That said, that if turbulent energy is injected at multiple scales, the density slope is shown to be particularly influenced by large-scale drivers, which make it difficult to trace the injection scale turnover \citep{YooCho14}. In addition, our power spectra provide critical information on the statistical properties of density fluctuations in three dimensions. The power spectrum of a column density or integrated emission image can be used to infer the power spectrum of the 3D density field when the tracer is in the ``thick slice'' limit. The ``thick slice'' limit is formally defined by \citet{lazPogo00} as the case where the velocity dispersion of the tracer is smaller than the channel width over which emission is integrated. In this case, velocity fluctuations are averaged out, and the measured 2D power spectrum can be directly related to the 3D density power spectrum. 
For the $\hi$, $^{12}$CO(1--0), and ${}^{13}$CO(1--0) in and around Perseus, this is indeed the case: the measured velocity dispersions ($\sim$1--2 km s$^{-1}$ for $^{12}$CO(1--0) and ${}^{13}$CO(1--0) and $\sim$5--7 km s$^{-1}$ for $\hi$) are smaller than the velocity range ($\delta$v = 20 km s$^{-1}$) over which each emission is summed. The same argument cannot be made for the 2MASS $A_{V}$ though, since it lacks velocity information. However, considering that the dust column density traces the total gas along a line of sight (atomic + molecular gas), we expect that the 2MASS $A_{V}$ is also most likely in the  ``thick slice'' limit. 

Whether or not velocity fluctuations are sufficiently averaged out can be more thoroughly examined by the Velocity Channel Analysis (VCA) developed by \citet{lazPogo00}. For this analysis, the velocity resolution of a data cube is varied from its instrumental resolution to coarser and coarser channel widths until the final integrated intensity image is created. At each velocity resolution, 2D power spectra are calculated and averaged together to estimate a representative slope. 
The velocity slices are then considered ``thick'' if the average slope saturates over several values of the velocity thickness. As a preliminary analysis, we applied the VCA to the $\hi$, $^{12}$CO(1--0), and ${}^{13}$CO(1--0) cubes of Perseus, finding that the power spectrum slope saturates well before the final channel width of 20 km s$^{-1}$. 
This suggests that we are indeed in the ``thick slice'' limit, probing density fluctuations in our 2D power spectrum analysis. While the VCA enables us to probe 3D velocity fluctuations which are potentially important for disentangling various drivers of turbulence (e.g., \citealt{OffnerArnce15}), it is out of the scope of this paper, since we mainly focus on comparing the statistical properties of density fluctuations revealed by several tracers. We plan to discuss the impact of velocity fluctuations through the VCA method in a forthcoming paper. 

\section{Discussion} 
\label{sec:discussion}

In this section, we discuss the results of the individual tracers and compare them with previous studies.
Several implications of our results will be also discussed.  

\subsection{Theoretical Predictions for the SPS of Turbulence}
\label{subsec:theory}

We begin our discussion by first summarizing theoretical expectations for the SPS of the multi-phase ISM (Table \ref{tab:TheoreticalPreds}). In the presence of supersonic turbulence without self-gravity, which is likely the case for the CNM in and around molecular clouds, the density spectral slope is expected to be shallower than the Kolmogorov index of $-11/3$ due to shock compression in three dimensions. Shocks can create small-scale density enhancements (e.g., \citealt{Beresnyak05, KowalLazarian07,Burkhart2010}), which in turn induce more power on small scales as compared to incompressible turbulence. In a weakly magnetized incompressible medium or a medium with no shock, the spectrum is very close to the Kolmogorov index of $-11/3$. Increasing the strength of the magnetic field in this limit increases the magnetic pressure and can steepen the power spectrum to $-13/3$ (e.g., \citealt{Kowal07}).

If self-gravity plays a role in shaping density structures, which is expected to be the case for the dense molecular medium, the 3D density and column density spectral slopes can become increasingly shallow as compared to non-gravitating supersonic turbulence (e.g., \citealt{Ossenkopf01, Collins12,FederrathKlessen13,BurkhartCollins15}). This is due to gravity enhancing over-densities in a supersonic flow. For example, \citet{BurkhartCollins15} examined the observational diagnostics of self-gravity by employing MHD simulations and found that self-gravitating supersonic turbulence can produce density structures that drive the spectral slope even up to positive values. This is in contrast to non-gravitating turbulence where the power is dominated by large-scale structures and decreases toward smaller scales. 

One of the important caveats for the above predictions is that the optical depth of the medium can significantly alter the column density/integrated intensity power spectral slope. For example, \citet{lazPogo04} and \cite{Burkhart13_a} showed that the integrated intensity images of the optically thick medium will have a power spectrum whose slope saturates to a universal value of $-3$ regardless of the presence of shocks, gravity, or magnetic fields. This would likely apply to ${}^{12}$CO(1--0), which is typically optically thick in molecular clouds, and possibly also to ${}^{13}$CO(1--0) and the cold $\hi$. 

\begin{table*}
\centering
\resizebox{\textwidth}{!}
{
\begin{tabular}{c c c} \hline \hline
Environment & 3D Density Spectrum$^{\rm a}$ & Reference \\ 
\tableline
$\mathcal{M}_s$ $\lesssim$ 1 & $\approx k^{-11/3}$ &  \citet{Kolmogorov1941, Goldreich95, ChoLazarian02,ChoLazarian03} \\
$\mathcal{M}_s$ $>$ 1 & shallower than $k^{-11/3}$& \citet{Beresnyak05, Kowal07} \\
$\mathcal{M}_s$ $>$ 1 and self-gravitating & shallower than pure compressible (positive values possible) & \citet{Fleck96, FederrathKlessen13, Collins12, BurkhartCollins15} \\
optically thick & $\approx k^{-3}$ & \citet{lazPogo04,Burkhart13_a};  \\
\hline
\end{tabular}}
{$^{a}$ For incompressible turbulence, the Kolmogorov power spectrum is $k^{-11/3}$, $k^{-8/3}$, and $k^{-5/3}$ for 3D, 2D, and 1D respectively.}
\caption{\label{tab:TheoreticalPreds} Power spectral slopes of turbulence for various environments}
\end{table*}

\subsection{$\hi$}
\label{subsec:hi_discussion}
Theoretical models of heating and cooling in the ISM predict the presence of two main atomic phases, the CNM and WNM (e.g., \citealt{Field1969, McKeeOst1977}). With distinct physical properties (density $n$ and kinetic temperature $T_{\rm k}$; $n$ $\sim$ 5--120 cm$^{-3}$ and $T_{\rm k}$ $\sim$ 40--200 K for the CNM; $n$ $\sim$ 0.03--1.3 cm$^{-3}$ and $T_{\rm k}$ $\sim$ 4100--8800 K for the WNM; e.g., \citealt{Wolfire03}), the CNM and WNM would then contribute to the overall turbulent properties of $\hi$ in different ways. Recently, \citet{Burkhart15_PDF} attempted to disentangle the turbulent properties of the CNM and WNM in Perseus by estimating $\mathcal{M}_s$ using two different constraints: (1) the $\hi$ spin temperatures derived by \citet{Stan14} based on the Arecibo $\hi$ emission and absorption observations and (2) the width of the $\hi$ column density PDF. While the $\mathcal{M}_s$ values based on the $\hi$ absorption data range from $\sim$1 to $\sim$40 with a median of $\sim$4, the average $\mathcal{M}_s$ deduced from the $\hi$ PDF width is $\sim$1. The authors then suggested that this discrepancy most likely arises from the fact that the $\hi$ absorption data mainly trace the CNM, while the $\hi$ PDF comes from a mix of the CNM and WNM. They also advocated that the $\hi$-to-H$_2$ transition could also contribute in narrowing the PDF (see also \citet{BialyBurkhartSternberg17}). Considering that our $\hi$ column density ($N$($\hi)$) traces both the CNM and WNM and the CNM fraction is relatively low in Perseus ($\sim$30\%; \citealt{Stan14}, we conclude that our derived power spectral slope of $-3.23\pm0.05$ is consistent with a non-gravitating transonic medium on average.

Given that the CNM fraction in and around Perseus is only $\sim$30\%, it is not surprising to find that the power spectral slope derived using the opacity-corrected data ($-3.22 \pm 0.05$) agrees well with the uncorrected case. However, at the same time, the opacity may impact the measurement of the power spectral slope for molecular clouds with higher CNM fractions, e.g., W43, where the opacity correction increases the $\hi$ mass by a factor of two \citep{Motte14, Bihr15}. 
 
\subsection{Dust} 
\label{subsec:dust_discussion}
Among the four tracers under study, the 2MASS $A_{V}$ shows the shallowest slope of $-2.72 \pm 0.12$, much shallower than the 3D Kolmogorov density slope of $-11/3$. Based on various numerical models of interstellar turbulence (Table ~\ref{tab:TheoreticalPreds}), this implies that the medium traced by the 2MASS $A_{V}$ is on average self-gravitating and supersonic. For example, \citet{BurkhartCollins15} examined how the slope of a column density power spectrum changes as a function of time by running simulations of non-gravitating and self-gravitating MHD turbulence with varying sonic and Alfv\'enic Mach numbers ($\mathcal{M}_{\rm s}$ and $\mathcal{M}_{\rm A}$) and found that the slope drastically changes as the model cloud evolves. The slope is initially $-11/3$ and becomes increasingly shallow for supersonic turbulence. Once gravity turns on, the slope increases beyond the purely supersonic case and eventually becomes positive. In their simulations, the slope values between $-2.7$ and $-2.5$ indeed appear during the self-gravitating supersonic phase (e.g., Figure 10 of \citealt{BurkhartCollins15}). Several independent studies of $A_{V}$ PDF also support our conclusion: a mix of log-normal and power-law shapes has been found for Perseus, which has been mainly interpreted as the presence of both supersonic turbulence and gravity (e.g., \citealt{Kainulainen11,BurkhartCollins15, Stanchev15}). 

In our analysis, we essentially use the 2MASS $A_{V}$ as a tracer of total hydrogen abundance based on the assumption that interstellar dust and gas are well mixed. This mixing of dust and gas has been inferred from the strong correlation between the color excess and the total hydrogen column density, $E(B-V)$/($N$($\hi$) + 2$N$(H$_{2}$)) = 1.7 $\times$ 10$^{-22}$ mag cm$^{2}$ (e.g., \citealt{Bohlin1978, Rachford09}). With the total-to-selective extinction ratio $R_{V}$ = 3.1, the typical value for the diffuse ISM (e.g., Mathis 1990), the correlation becomes $A_{V}$/($N$($\hi$) + 2$N$(H$_{2}$)) = 5.3 $\times$ 10$^{-22}$ mag cm$^{2}$. This dust-to-gas ratio, however, changes in different environments, e.g., $R_{V}$ is higher with $\sim$4--6 in dense molecular clouds, resulting in higher dust-to-gas values (e.g., \citealt{Cardelli98, Fitzpatrick99}). To assess the impact of the dust-to-gas ratio variation on our power spectral slope, we then perform a simple test by taking any pixel with $A_{V}$ $\geq$ 5 mag and replacing it with the half of the original value, e.g., the pixels with $A_{V}$ = 6 mag are replaced with 3 mag. This test is to simulate the case where the dust-to-gas ratio increases in dense regions by a factor of two, and we find that the power spectral slope of the simulated $A_{V}$ image is almost the same as before, $-2.72\pm0.15$, suggesting that our interpretation of the self-gravitating and supersonic medium traced by the 2MASS $A_{V}$ is likely robust. 

Another interesting thing to note is that the 2MASS $A_{V}$, which traces both the atomic and molecular media, shows the slope that is distinctly different from the $\hi$ slope. In terms of mass, the neutral medium of Perseus is dominated by $\hi$ ($\hi$ and H$_{2}$ mass of $\sim$2 $\times$ 10$^{4}$ M$_{\odot}$ and $\sim$6 $\times$ 10$^{3}$ M$_{\odot}$ respectively; calculated over the $\hi$ coverage in Figure \ref{fig:data}), and a naive expectation would then be that the 2MASS $A_{V}$ slope is similar to the $\hi$ slope. Our finding of the shallower slope for $A_{V}$ hence suggests a strong influence of gravity, which has in fact been noted in several numerical studies of interstellar turbulence. For example, \citet{BurkhartCollins15} showed that self-gravitating high-density cores on small scales have a significant impact on the global turbulence statistics of the model cloud (e.g., column density PDF and the SPS) and found that the introduction of a delta-function like density profile on small scales can mimic self-gravity in the turbulence power spectrum. 

\subsection{${}^{12}$CO(1--0) and ${}^{13}$CO(1--0)}
\label{s:12CO_discussion}
As for $^{12}$CO, we find a relatively steep slope of $-3.08\pm0.08$, which is consistent with $-3$, implying that the $^{12}$CO(1--0) emission in Perseus is likely subject to the optical depth effect. The presence of self-absorbing medium makes a substantial impact on the slope of an intensity power spectrum, as shown by several authors including \citet{lazPogo04} and \cite{Burkhart13_a}. In particular, \citet{lazPogo04} predicted that absorption induces a universal slope of $-3$ for intensity fluctuations, and \cite{Burkhart13_a} numerically confirmed this prediction by analyzing $^{13}$CO(2--1) power spectra of MHD simulations with varying sonic and Alfv\'{e}nic Mach numbers. 
In the case of optically thick emission with $\tau \gg 1$, \cite{Burkhart13_a} found that the integrated intensity spectral slope saturates to $-3$ regardless of $\mathcal{M}_{\rm s}$ and $\mathcal{M}_{\rm A}$ values. On the other hand, for the mildly optically thick medium with $\tau \sim 1$, the behavior was mixed: super-Alfv\'enic turbulence shows an increasingly shallow slope as $\mathcal{M}_{\rm s}$ increases (from $-3$ to $-2$ for $\mathcal{M}_{\rm s}$ $\sim$ 0.4--8), while the slope is always around $-3$ for sub-Alfv\'enic turbulence. For Perseus, we find that $^{12}$CO is somewhat optically thick with $\tau$($^{12}$CO) $\sim$ 0.3--1 across the cloud, based on the curve-of-growth analysis of $^{13}$CO by \citet{Pineda08} (Table \ref{tab:sub_regions}). These estimates are, however, for individual dark and star-forming regions on $\sim$10 pc scales: $\tau$($^{12}$CO) will be most likely $\gg$ 1 for smaller and denser cores on $\sim$0.07 pc scales ($^{12}$CO pixel size). Indeed, \citet{Pineda08} found that $I$($^{12}$CO) becomes saturated at $A_{V}$ $\gtrsim$ 4 mag in Perseus on $\sim$0.4 pc scales (2MASS $A_{V}$ pixel size), indicating that the emission becomes optically thick. The same argument applies for $^{13}$CO. While the measured slope of $-2.88\pm0.07$ is not as consistent with $-3$ as it is for $^{12}$CO, $\tau$($^{13}$CO) is in fact not drastically different from $\tau$($^{12}$CO) (Table \ref{tab:sub_regions}): only two regions, B5 and Westend, show $\tau$($^{12}$CO) more than a factor of two higher than $\tau$($^{13}$CO). In addition, the $^{13}$CO(1--0) emission was found to saturate at $A_{V}$ $\gtrsim$ 5 mag \citep{Pineda08}. Finally, we note that $^{12}$CO(3--2) and $^{13}$CO(2--1) are also likely subject to the optical depth effect, based on \citet{Sun06} who measured the slopes of $-3.15\pm0.04$ and $-3.03\pm0.14$ for the two transitions. 

Other than the opacity, the $^{12}$CO(1--0) and $^{13}$CO(1--0) emission could also be subject to the effects of CO chemistry, e.g., formation (CO abundance sharply drops in diffuse regions with $A_{V}$ $\lesssim$ 1--2 mag due to insufficient dust shielding; e.g., \citealt{Wolfire10}) and depletion (CO is frozen onto dust grains in dense regions with $A_{V}$ $\gtrsim$ 5--10 mag; e.g., \citealt{Bergin98}). To evaluate the impact of CO chemistry on our power spectrum analysis, we then perform a similar test on the 2MASS $A_{V}$ image as we do for the dust-to-gas ratio variation in Section \ref{subsec:dust_discussion}. Specifically speaking, we replace any pixels with $A_{V}$ $<$ 2 mag with zero values and re-measure the power spectral slope to mimic the effect of CO formation. Similarly, all pixels with $A_{V}$ $>$ 5 mag are fixed to 5 mag to simulate CO depletion in dense regions. In the CO depletion experiment, the power spectral slope of $-2.77\pm0.13$ is comparable to the original value of $-2.72\pm0.12$. In addition, we found that the new slope of $-2.39\pm0.07$ for the CO formation test is reasonably close to the original value, implying that CO formation and depletion likely do not significantly affect our SPS results. While this conclusion is consistent with \citet{Padoan06_b}, who performed essentially the same tests on Taurus, another low-mass star-forming region, we note that more molecular clouds with diverse properties should be further examined to confirm the conclusion. All in all, our results suggest that particular attention is needed to use $^{12}$CO(1--0) and $^{13}$CO(1--0) as a probe of the turbulent properties of molecular gas, since the emission may suffer from the effects of opacity and chemistry. 
 
\begin{table}
\centering
\caption{\label{tab:sub_regions} Properties of the dark and star-forming regions$^{\rm a}$ in Perseus}
\begin{tabular}{l c c c c c} \hline \hline 
Property & B5 & IC 348 & B1 & NGC 1333 & Westend \\ \hline 
$^{\rm b}$$\tau$($^{12}$CO) & 1.0 & 0.5 & 0.5 & 0.3 & 0.9 \\
$^{\rm c}$$\tau$($^{13}$CO) & 0.3 & 0.4 & 0.3 & 0.3 & 0.4 \\ 
\hline 
\end{tabular}
\\
{$^{\rm a}$ These sub-regions are labeled in Figure \ref{fig:data}. \\
$^{\rm b}$ These values are derived using Equation (14) of \citet{Pineda08} along with their Tables 2 and 3. \\
$^{\rm c}$ From Table 2 of \citet{Pineda08}.} 
\end{table}

\subsection{Comparison to Previous Studies}
\label{s:final_discussion}
In this section, we make a comparison with previous SPS studies to place our results into a larger context. For the comparison, we consider density fluctuations in various Galactic environments (e.g., non-star-forming clouds and CNM-dominated regions) and focus on $\hi$ and dust power spectra due to potential opacity and chemical effects with the molecular gas tracers. The details on the previous studies, including the measured SPS slopes and spatial scales, are presented in Table \ref{tab:previous_studies}.

First of all, we find that our results are generally consistent with the previous studies for the Milky Way: the $\hi$ SPS shows a Kolmogorov-like slope (from $-4$ to $-3.6$), 
while the dust SPS has a much shallower slope (from $-2.9$ to $-2.7$). This finding suggests that on average, $\hi$ emission traces a non-gravitating transonic/subsonic medium, while dust probes a self-gravitating supersonic medium. The only exceptions are \citet{Deshpande00} and \citet{Martin15}, where the $\hi$ SPS slopes were found to be shallower than $-3$. In the case of \citet{Deshpande00}, their data strictly trace the CNM seen in absorption against Cassiopeia A and Cygnus A, and the measured slope of $-2.75\pm0.25$ in fact agrees with the expectation for the supersonic CNM. On the other hand, the analysis by \citet{Martin15} for intermediate Galactic latitudes is based on $\hi$ emission observations (tracing both the CNM and WNM), and whether their shallow slopes of $\gtrsim$ $-3$ result from a substantial amount of the CNM remains to be further examined. 

Interestingly, no break has yet been observed in the SPS analyses probing the spatial scales from $\sim$0.01 pc to a few tens pc. One possible explanation is that interstellar turbulence is driven on the scales larger than individual molecular clouds. Possible turbulence drivers then include galactic-scale thermal, gravitational, and magnetorotational instabilities, accretion of circumgalatic material, as well as spiral shock waves (e.g., \citealt{DobbsBonnell07, Brunt09,krumBurk16}). Small-scale stellar feedback such as outflows and bubbles is likely not the dominant source for turbulence (e.g., \citealt{Nestingen-Palm17}), considering comparable SPS slopes between non-star-forming (e.g., Polaris flare, Ursa Major cirrus clouds, and MBM16) and star-forming regions over a wide range of length scales (e.g., Perseus and the SMC).

In particular, for Perseus, \citet{Padoan09} investigated both density and velocity fluctuations in NGC1333, an active star-forming and outflow-driving region in Perseus. If turbulence is primarily driven by outflows on the scale of individual stellar clusters, the power spectral slope should flatten beyond the energy injection scale (e.g., \citealt{Matzner07}), which \citet{Padoan09} estimated to be 0.3 pc for NGC1333. Our measured power spectrum for $^{13}$CO(1--0) does not show any characteristic length scale. Finally, the clear power-laws down to $\sim$0.01 pc scales suggests that turbulence dissipation, likely via viscous dissipation and ambipolar diffusion (e.g., \citealt{HennebelleFalarone12}), occurs on the scales smaller than 0.01 pc.

Lastly, we consider how the SPS performs overall as a diagnostic tool for characterizing turbulence over other statistical techniques. Recently, \citet{Boyden16} evaluated several astrostatistics (e.g., Principal Component Analysis (PCA), spectral correlation function, SPS, etc.) as a probe of stellar wind feedback by applying these statistics to synthetic $^{12}$CO(1--0) cubes generated from MHD simulations. In their experiments, the SPS showed a strong response to time evolution, but ambiguous behavior to wind activities and very little response to magnetic fields (although due to opacity effects, $^{12}$CO(1--0) may not be an optimal tracer for such a study). In addition, by having two different scales of energy injection in MHD simulations, \cite{YooCho14} demonstrated that the small-scale driving could have only a limited influence on the density fluctuations.   
These results suggests that the SPS maybe not a good diagnostic tool to study small-scale stellar feedback such as stellar winds, and thus utilizing a suit of statistical tools whose individual strengths probe different observable characteristics would be necessary to fully characterize turbulence in the multi-phase ISM. 

\begin{table*}
\begin{center}
\caption{\label{tab:previous_studies} Summary of Previous SPS Studies}
\begin{tabular}{lccccc} \hline \hline
Object & Tracer & Optical Depth & Scale & Slope & Reference \\ \hline
Polaris Flare & Dust & Thin & 0.01$\sim$8 pc & $-2.7\pm0.1$ & (1) \\
High Latitude Cirrus & Dust & Thin & 0.01$\sim$50 pc & $-2.9\pm0.1$ & (2) \\
CNM & $\hi$ Absorption & Thin & 0.01$\sim$3 pc & $-2.75\pm0.25$ & (3) \\
WNM & $\hi$ Emission & Thin & -- & $-4$ & (4) \\
Ursa Major & $\hi$ Emission & Thin & 0.1$\sim$25 pc & $-3.6\pm0.2$ & (5) \\
MBM16 & $\hi$ Emission & Thin & 0.1$\sim$20 pc & $-3.7\pm0.2$ & (6) \\ 
\multirow{2}{*}{North Ecliptic Pole} & \multirow{2}{*}{$\hi$ Emission} & \multirow{2}{*}{Thin} & \multirow{2}{*}{--} & from $-2.86\pm0.04$ & \multirow{2}{*}{(7)} \\
                                     &                                 &                       &                     & to $-2.59\pm0.07$ & \\ 
\hline
\end{tabular}
\end{center}
{Reference: \\
(1) \citet{Miville10}: \textit{Herschel} observations at 250 $\mu$m, 350 $\mu$m, and 500 $\mu$m were used to study the diffuse interstellar cloud Polaris flare. \\ 
(2) \citet{Miville16}: Dust-scattered light in the CFHT $g$-band was analyzed. \\ 
(3) \citet{Deshpande00}: The cold HI seen in absorption against Cassiopeia A and Cygnus A was examined. \\ 
(4) \citet{Dickey01}: One patch of the sky observed in the Southern Galactic Plane Survey was probed. \\ 
(5) \citet{Miville03}: The non-star-forming, high latitude cirrus cloud Ursa Major was under study. \\ 
(6) \citet{Pingel13} \\
(7) \citet{Martin15}: GBT HI emission observations of low-, intermediate-, and high-velocity clouds at the north ecliptic pole were analyzed.}
\end{table*}

\section{Conclusion and future work}
\label{sec:conclusion}
In this study we presented a comparison of the two dimensional spatial power spectrum (SPS) derived from four distinct ISM tracers for the Perseus molecular cloud. In our derivation of the SPS, we take median value of the pixel distribution in each annulus as opposed to the standard practice of taking the mean to mitigate the effects of the Gibbs phenomenon. We find the $\hi$ data produce the steepest slopes with values of $-3.23\pm0.05$ and $-3.22\pm0.05$ for the uncorrected and corrected $\hi$, respectively. The dust gives the shallowest slope values of $-2.72\pm0.12$. In the case of ${}^{12}$CO(1--0) and ${}^{13}$CO(1--0) we measure slopes of $-3.08\pm0.08$ and $-2.88\pm0.07$, respectively. The comparison between the relative slopes of the tracers reveals several important characteristics about the turbulent environment in Perseus: 

\begin{itemize}

\item Our derived slope value for the $\hi$ in Perseus suggests that the $\hi$ is largely non-gravitating and, in general, transonic. The consistency of the slope between the opacity corrected and uncorrected column density images is not surprising since the CNM fraction in Perseus is only about 30\%. This may not be the case for molecular clouds with significant CNM fractions.

\item We assume the dust dependably traces the gas in Perseus. It is therefore interesting to see the dust have such a shallow slope as compared to the $\hi$ (since $\hi$ dominates the mass of Perseus). Simple tests which increases the dust-to-gas ratio in dense regions of the dust image reveal a similar slope value, which according to numerical studies, indicates the dust in Perseus is influenced by self-gravity due to the relatively shallow slope. 

\item While the dust is rather robust to small-scale dust-to-gas variations, we find the ${}^{12}$CO(1-0) and ${}^{13}$CO(1--0) power spectra are particularly  susceptible to opacity effects --- confirming theoretical prediction of a slope of $\sim$$-$3 when the SPS is applied to an optically thick medium. Furthermore, we show through simple models of CO formation/depletion that such chemical processes do not significant affect our power spectrum analysis for Perseus specifically. This conclusion may not be applicable for other molecular clouds with more complicated chemistry. We conclude careful consideration is needed when utilizing ${}^{12}$CO(1-0) and ${}^{13}$CO(1-0) intensity maps as a probe for turbulent environments due to possible opacity and chemical effects. 

\item Previous results of dust power spectra in diffuse, high-latitude clouds agree well with our dust slope. The similar slope values measured between regions of active star formation (i.e., Perseus) and non-star forming (Polaris Flare and Ursa Major cirrus clouds) hint that small-scale stellar feedback is relatively unimportant in regulating the turbulent environment traced by interstellar dust. A comparison of previous $\hi$ results for the SPS derived over a diverse set of environments (e.g. star forming vs. non-star forming and CNM dominated regions) shows our $\hi$ slope is in good general agreement, and is indicative that $\hi$ traces a non-gravitating transonic to supersonic medium in Galactic environments. 

\item The lack of a break in any of the power spectra signifies we detect neither the injection nor the dissipation scale in Perseus for any of our tracer. Additionally, the absence of a break in any previous Galactic SPS analysis suggests large-scale turbulence drivers, such as galactic-scale instabilities driven by gravitational, magnetorotational or supernovae dominate over small-scale drivers like stellar outflows. 

\end{itemize}

\acknowledgments

\noindent B.B.~is supported by the NASA Einstein Postdoctoral Fellowship and the ITC Postdoctoral Fellowship at the Harvard-Smithsonian Center for Astrophysics. 

\bibliographystyle{apj}
\bibliography{apj-jour,MasterRefs}

\end{document}